\documentclass[sigconf,nonacm]{acmart}

\setcopyright{none}
\renewcommand\footnotetextcopyrightpermission[1]{}
\usepackage{xspace}
\usepackage{graphicx}
\usepackage{subcaption}
\usepackage{enumitem}
\usepackage{booktabs}
\usepackage{multirow}
\usepackage[ruled,linesnumbered,vlined]{algorithm2e}
\usepackage{tikz}
\usepackage{xcolor}

\newcommand*\circled[1]{\tikz[baseline=(myanchor.base)] \node[circle,fill=.,inner sep=1pt] (myanchor) {\color{-.}\bfseries\footnotesize #1};}

\def\Snospace~{\S{}}

\newcommand{\sys}{Curator\xspace}

\newcommand{\paraspace}{\vspace{0.05in}}
\newcommand{\parab}[1]{\paraspace\noindent{\bf #1} }

\begin{document}
\title{\sys: Efficient Vector Search with Low-Selectivity Filters}
\author{Yicheng Jin}
\affiliation{%
  \institution{Duke University}
  \city{Durham}
  \state{NC}
  \country{USA}
}
\email{yicheng.jin@duke.edu}
\author{Yongji Wu}
\affiliation{%
  \institution{University of California, Berkeley$^1$}
  \city{Berkeley}
  \state{CA}
  \country{USA}
}
\email{wuyongji317@gmail.com}
\author{Wenjun Hu}
\affiliation{%
  \institution{Yale University}
  \city{New Haven}
  \state{CT}
  \country{USA}
}
\email{wenjun.hu@cantab.net}
\author{Bruce M. Maggs}
\affiliation{%
  \institution{Duke University}
  \city{Durham}
  \state{NC}
  \country{USA}
}
\email{bmm@cs.duke.edu}
\author{Jun Yang}
\affiliation{%
  \institution{Duke University}
  \city{Durham}
  \state{NC}
  \country{USA}
}
\email{junyang@cs.duke.edu}
\author{Xiao Zhang}
\affiliation{%
  \institution{Cisco ThousandEyes$^1$}
  \city{Durham}
  \state{NC}
  \country{USA}
}
\email{xz234@alumni.duke.edu}
\author{Danyang Zhuo}
\affiliation{%
  \institution{Duke University}
  \city{Durham}
  \state{NC}
  \country{USA}
}
\email{danyang@cs.duke.edu}

\begin{abstract}
Embedding-based dense retrieval has become the cornerstone of many critical applications, where approximate nearest neighbor search (ANNS) queries are often combined with filters on labels such as dates and price ranges. Graph-based indexes achieve state-of-the-art performance on unfiltered ANNS but encounter connectivity breakdown on low-selectivity filtered queries, where qualifying vectors become sparse and the graph structure among them fragments. Recent research proposes specialized graph indexes that address this issue by expanding graph degree, which incurs prohibitively high construction costs.
Given these inherent limitations of graph-based methods, we argue for a dual-index architecture and present \sys, a partition-based index that complements existing graph-based approaches for low-selectivity filtered ANNS. \sys builds specialized indexes for different labels within a shared clustering tree, where each index adapts to the distribution of its qualifying vectors to ensure efficient search while sharing structure to minimize memory overhead. The system also supports incremental updates and handles arbitrary complex predicates beyond single-label filters by efficiently constructing temporary indexes on the fly.
Our evaluation demonstrates that integrating \sys with state-of-the-art graph indexes reduces low-selectivity query latency by up to 20.9$\times$ compared to pre-filtering fallback, while increasing construction time and memory footprint by only 5.5\% and 4.3\%, respectively.
\end{abstract}

\maketitle

\setcounter{footnote}{1}
\footnotetext{Yongji Wu and Xiao Zhang were with Duke University at the time this work was conducted.}

\section{Introduction}
\label{sec:intro}

The rise of deep learning has transformed information retrieval by enabling the representation of unstructured data as high-dimensional vectors that capture semantic relationships~\cite{le2014distributed,radford2021learning,narayanan2017graph2vec}. Approximate nearest neighbor search (ANNS) over these vector embeddings has become a fundamental primitive powering modern applications from web search~\cite{Nayak_2019,Zeng_2022} to recommendation systems~\cite{zhang2019deep,naumov2019deep,covington2016deep} and retrieval-augmented generation~\cite{chatgpt_retrieval_plugin,lewis2020retrieval}.

Pure semantic similarity, however, often proves insufficient for real-world applications. Modern systems increasingly require \emph{filtered ANNS}, where vector similarity search is constrained by structured predicates over metadata attributes. For instance, e-commerce platforms must find semantically similar products within specific price ranges and categories, while enterprise search systems need to respect user access permissions and document types. This combination of semantic similarity and structured filtering has driven significant interest from both industry~\cite{wei2020analyticdb,pinecone,wang2021milvus,weaviate,qdrant} and research communities~\cite{wang2022navigable,wu2022hqann,gollapudi2023filtered,zuo2023arkgraph,zhang2023vbase,patel2024acorn,zuo2024serf,cai2024navigating}.

Among the various challenges in filtered ANNS, handling queries with diverse \emph{selectivity}---the fraction of vectors qualifying the filter conditions---is particularly difficult. The difficulty varies dramatically across the selectivity spectrum, with each regime requiring different algorithmic strategies and presenting distinct performance bottlenecks.

\subsection{Why Existing Approaches Fail at Low Selectivity}

\begin{figure}[t]
\centering
\includegraphics[width=\linewidth]{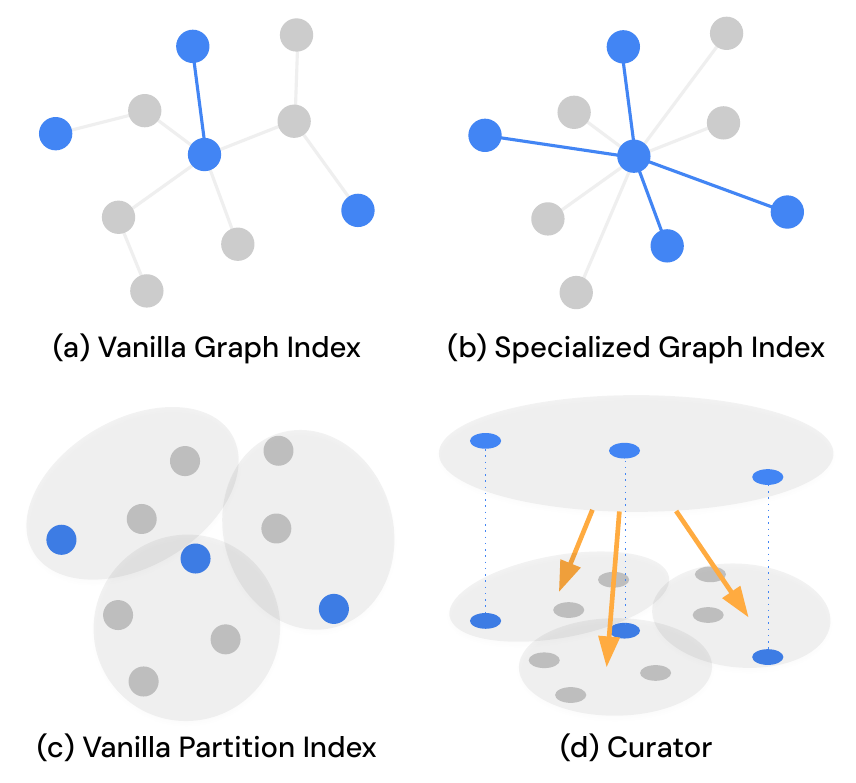}
\caption{Comparison of different approaches to filtered ANNS. Blue points represent qualifying vectors, gray points represent non-qualifying vectors. (a) Vanilla graph-based approach suffers from connectivity breakdown at low selectivity. (b) Specialized graph-based approach requires increasing graph degree significantly to maintain connectivity. (c) Vanilla partition-based approach suffers from granularity mismatch at low selectivity. (d) \sys adopts a hierarchical partitioning approach that buffers qualifying vectors in sparse regions at higher levels of the tree, adapting clustering granularity to filter selectivity.}
\label{fig:motivation}
\end{figure}

Existing approaches to filtered ANNS can be broadly categorized by their underlying index structures: flat indexes (brute-force search), graph-based indexes, and partition-based indexes. While each index type has found success in specific scenarios, they all encounter limitations when handling low-selectivity queries. Fig.~\ref{fig:motivation} illustrates these challenges across different approaches.

\parab{Graph connectivity breakdown.} As shown in Fig.~\ref{fig:motivation}(a), filtered search on graph-based indexes exposes a critical trade-off: as selectivity decreases, the sub-graph of qualifying vectors becomes increasingly sparse and disconnected. This connectivity breakdown forces algorithms to choose between two unsatisfactory options: visit all neighbors of each node, which is expensive since most neighbors are unqualified, or skip unqualified neighbors entirely, which is both incomplete—as some qualifying vectors become unreachable—and inefficient, since graph search requires sufficient connectivity for optimal performance.

\parab{Densification becomes prohibitive.} Recent approaches attempt to address connectivity issues through graph densification, as illustrated in Fig.~\ref{fig:motivation}(b). Both ACORN~\cite{patel2024acorn} and Filtered DiskANN~\cite{gollapudi2023filtered} build dense graphs to ensure connectivity among qualifying vectors. However, the required graph degree grows rapidly as selectivity decreases. If each node needs $m$ qualified neighbors and selectivity is $s$, then the total degree must be approximately $m/s$. This relationship makes densification prohibitively expensive for low-selectivity queries, as our empirical evaluation demonstrates significant increases in construction time and memory usage. Alternative approaches like UNG~\cite{cai2024navigating} attempt different strategies but face significant limitations, as detailed in \autoref{sec:background}.

\parab{Pre-filtering scales poorly.} When densification becomes impractical, many systems fall back to pre-filtering: first identifying all qualifying vectors, then performing brute-force search over this subset. While this approach guarantees completeness, it scales poorly with the absolute number of qualifying vectors. On large-scale datasets, low-selectivity filters can still match tens of thousands of vectors, making brute-force search computationally expensive.

\parab{Partition granularity mismatch.} Given these limitations of graph-based approaches, one might consider partition-based alternatives. Indexes like IVF (Inverted File) cluster vectors and reduce search scope by examining only nearby clusters, and they are not vulnerable to connectivity breakdown. However, naively applying them incurs another issue: partitions built for all indexed vectors are too fine-grained for filtered search. As shown in Fig.~\ref{fig:motivation}(c), when selectivity is low, each cluster contains only a few qualifying vectors, thus the search must visit significantly more clusters compared to unfiltered search, leading to substantial overhead for distance computations against cluster centroids.

\subsection{Our Approach: Adaptive Hierarchical Partitioning}

\begin{figure}[t]
\centering
\includegraphics[width=\linewidth]{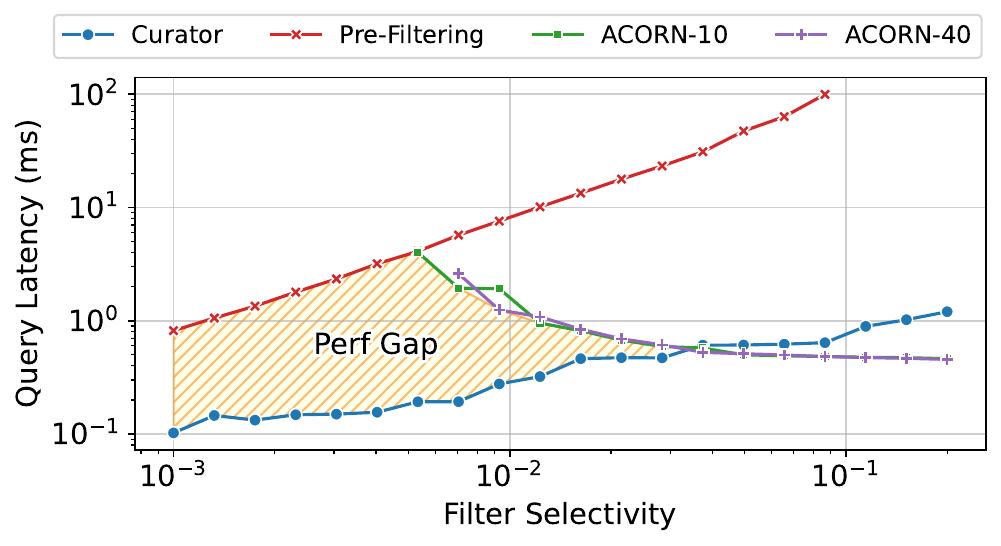}
\caption{Performance comparison between \sys and the state-of-the-art solution (ACORN + pre-filtering) across varied selectivity levels in the low-selectivity regime. ACORN-$\gamma$ ($\gamma = 10, 40$) denotes ACORN with different graph degrees, with larger $\gamma$ indicating denser index graphs. \sys closes the performance gap of ACORN at low selectivity with minimal overhead. Detailed experiment setup is provided in \autoref{sec:eval:sys_perf}.}
\label{fig:sys_perf}
\end{figure}

The limitations across existing approaches suggest the need for a specialized index dedicated to low-selectivity filters. This raises a natural question: should we design a dedicated index or improve existing specialized indexes to handle a wider range of selectivity? Given the fundamental connectivity challenges that graph-based indexes face at low selectivity, we design \sys to complement existing graph-based approaches through a dual-index architecture rather than attempting to extend their performance into the low-selectivity regime. Graph-based methods excel at high-selectivity scenarios, while \sys specializes in the low-selectivity regime where traditional approaches struggle. This complementary design enables the two indexes to work together, achieving better overall performance across the entire selectivity spectrum. The key requirement for this approach to be cost-effective is that \sys must maintain low memory and construction overhead.

Following this complementary design philosophy, this paper introduces \sys, a partition-based index that employs adaptive hierarchical clustering to achieve efficient low-selectivity search. As a partition-based approach, \sys deliberately accepts slower search performance than graph-based indexes at high selectivity, where graph connectivity remains strong, in exchange for robust performance at low selectivity where graph-based approaches suffer from connectivity breakdown. As illustrated in Fig.~\ref{fig:motivation}(d), \sys uses hierarchical partitioning where qualifying vectors in sparse regions are buffered at higher levels of the tree, adapting clustering granularity to the local density of qualifying vectors. The system constructs specialized indexes for different filter predicates, all embedded within a shared hierarchical clustering tree that maintains low overhead through structural sharing.

\sys supports both label-based filtering—where vectors are associated with categorical labels known in advance and queries return the nearest vectors matching a given label—and arbitrary complex predicates involving multiple attribute types and logical operators. The core challenge with complex predicates is that they are typically not indexed directly due to the vast number of possible combinations. Prior works resort to inefficient multi-stage filtering or inflexible special handling of a small subset of predicate types in index design. \sys addresses this by dynamically constructing temporary indexes that mirror the shared tree structure with minimal overhead, enabling efficient evaluation of complex predicates without requiring pre-built specialized indexes for every combination.

To validate this complementary design, we evaluate \sys against ACORN (the state-of-the-art graph-based index) using a semi-synthetic dataset constructed from YFCC-10M with 20 logarithmically distributed selectivity levels in $[0.001, 0.2]$, enabling systematic coverage of the low-selectivity regime. As shown in Fig.~\ref{fig:sys_perf}, brute-force search (pre-filtering fallback) exhibits linear scaling, while ACORN shows sharp performance degradation at low selectivity. \sys fills this gap with minimal construction and memory overhead. The results show that increasing graph degree (from ACORN-10 to ACORN-40) offers limited performance improvement at low selectivity, confirming that densifying graphs is not an effective solution and making \sys an ideal complementary approach.

\subsection{Contributions}

This paper makes the following key contributions:

\begin{itemize}
    \item \textbf{Adaptive partitioning for low-selectivity ANNS}: We design \sys, a hierarchical partition-based index that adapts clustering granularity to the distribution of qualified vectors, achieving high performance for low-selectivity queries while maintaining minimal memory overhead.
    
    \item \textbf{Efficient complex predicate support}: We present a novel algorithm for handling arbitrary predicates by dynamically constructing temporary indexes with minimal overhead.
    
    \item \textbf{Comprehensive system design}: We develop efficient algorithms including batch construction, incremental updates, and specialized search procedures for both single-label filters and complex predicates, ensuring search efficiency and completeness while adapting to dynamic changes in data distributions.
    
    \item \textbf{Extensive experimental validation}: We conduct comprehensive experiments on real-world datasets demonstrating that \sys reduces low-selectivity query latency by up to 20.9$\times$ compared to the state-of-the-art solution (ACORN + pre-filtering), while increasing construction time and memory footprint by only 5.5\% and 4.3\%, respectively, confirming the effectiveness of our complementary approach.
\end{itemize}
\section{Background}
\label{sec:background}

\subsection{Problem Formulation}

We formulate the problem of filtered ANNS as follows. We start with single-label search and then generalize to complex predicate search:

\parab{Single-label search.} Let $\mathcal{L} = \{l_1, l_2, \ldots, l_{|\mathcal{L}|}\}$ represent a finite universe of \textit{labels}, and consider a dataset $S = \{(x, L_x) \mid L_x \subseteq \mathcal{L}\}$ where each object consists of a vector $x$ and an associated set of labels $L_x$. Single-label search involves queries that check for the existence of a single label in the label set: a query $(x_q, l, k)$ requests the $k$ nearest neighbors of the query vector $x_q$ within the subset of vectors where $l \in L_x$.

\parab{Complex predicate search.} Consider a more general dataset $S = \{(x, A_x)\}$ where each object consists of a vector $x$ and associated attributes $A_x$ including categorical labels, numerical values, timestamps, and other structured data. Complex predicate search involves arbitrary filter predicates $\sigma$ combining range queries, composite conditions across multiple attributes, and complex logical operators. A query $(x_q, \sigma, k)$ requests the $k$ nearest neighbors of the query vector $x_q$ within the subset $\mathcal{P}_{\sigma}$ of vectors qualifying predicate $\sigma$. Complex predicate search subsumes single-label search and additionally supports AND queries $(x_q, L_q, k)$ where $L_q \subseteq L_x$, and OR queries $(x_q, L_q, k)$ where $L_q \cap L_x \neq \emptyset$.

For both search types, the \textit{selectivity} of filter $\sigma$ is defined as $|\mathcal{P}_{\sigma}|/|S|$, which fundamentally determines query complexity. We measure search quality using $\texttt{Recall@K} = \frac{|R \cap R^*|}{k}$, where $R$ and $R^*$ denote the retrieved results and ground truth respectively.

\subsection{Filtered ANNS Approaches}

Below, we review existing methods of filtered ANNS organized by how filtering is integrated. This taxonomy is orthogonal to the underlying index type, focusing instead on how adaptations are applied to search algorithms and index structures. For each approach, we first discuss its application to single-label search scenarios and then examine its support for complex predicate search.

\parab{Metadata filtering.} This approach integrates filtering with no or minimal modifications to search algorithms and index structures, utilizing a single shared index. It includes three subtypes: (1) \emph{Pre-filtering}~\cite{weaviate,wang2021milvus,wei2020analyticdb} retrieves all qualified vectors from a scalar index and performs brute-force search, since most vector indexes cannot efficiently support searching subsets of vectors. (2) \emph{Post-filtering}~\cite{wang2021milvus,wei2020analyticdb} filters intermediate results from unfiltered ANNS, though determining sufficiently large $k$ to ensure enough results remain after filtering is challenging and overshoot leads to inefficiency. (3) \emph{Inline filtering}~\cite{pinecone,johnson2019faiss,hnswlib,zhang2023vbase} excludes unqualified vectors during search, eliminating intermediate $k$ selection but requiring explicit algorithmic support. This approach can support arbitrary complex predicates.

\parab{Per-label indexing.} This approach constructs dedicated vector indexes for each label and directs queries to the corresponding index without changing search algorithms~\cite{parlayivf}. While it delivers optimal search performance for single-label queries by eliminating query-time filtering, it incurs significant construction and memory overhead since each vector may be indexed multiple times across different per-label indexes. Extending this to per-predicate indexing for complex predicate search is generally impossible due to the combinatorial explosion of possible predicates.

\parab{Specialized indexes.} Recent research has developed specialized indexes for filtered ANNS modifying both index construction and search algorithms, with most work focusing on graph-based approaches. These methods can be categorized into three main strategies:

(1) \emph{Fused distance.} These methods~\cite{wang2022navigable,wu2022hqann} employ hybrid distance metrics that integrate both attribute and vector similarity throughout index construction and query processing. They are limited to equality-based attribute matching and require queries to specify values of all attributes for attribute similarity computation. This severely restricts query expressiveness, making them unsuitable for the problem settings we are considering.

(2) \emph{Graph densification.} Filtered DiskANN~\cite{gollapudi2023filtered} and ACORN~\cite{patel2024acorn} follow a two-phase approach: dense graph construction then compression. Filtered DiskANN builds dense graphs by merging per-label indexes, while ACORN directly builds large-degree graphs. For compression, Filtered DiskANN prunes edges while preserving connectivity, whereas ACORN removes edges between nodes reachable as two-hop neighbors, compensating by visiting both direct and two-hop neighbors during search. Both face the fundamental connectivity challenge that the required graph degree scales with $1/\text{selectivity}$. When this scaling becomes prohibitive, ACORN falls back to pre-filtering.

(3) \emph{Subgraph stitching.} UNG~\cite{cai2024navigating} constructs separate graph indexes for each unique label set and connects them using cross-group edges. During query processing, it traverses subgraphs corresponding to matching label sets, with connectivity among label sets guaranteed by cross-group edge construction to ensure completeness. However, this approach often becomes inefficient when label set groups become highly fragmented with few nodes per group, which frequently occurs in real datasets. In such cases, the graph structure degrades as most edges become cross-group connections that prioritize completeness over search efficiency. This leads to significant regression in search performance and long construction times for the stitching process in our evaluation.

These specialized approaches exhibit varying support for complex predicates due to their tight integration of filtering with index structures. Filtered DiskANN supports only OR queries, and UNG supports only AND queries, while ACORN extends beyond these limitations to handle general complex predicates.

\section{\sys Overview}
\label{sec:overview}

\begin{figure}[t]
  \centering
  \includegraphics[width=0.85\columnwidth]{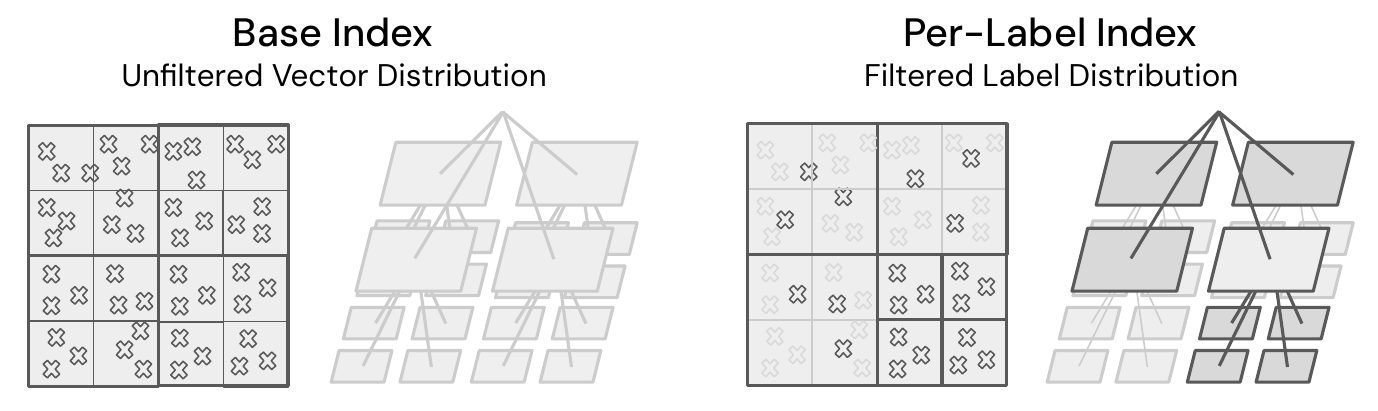}
  \includegraphics[width=0.85\columnwidth]{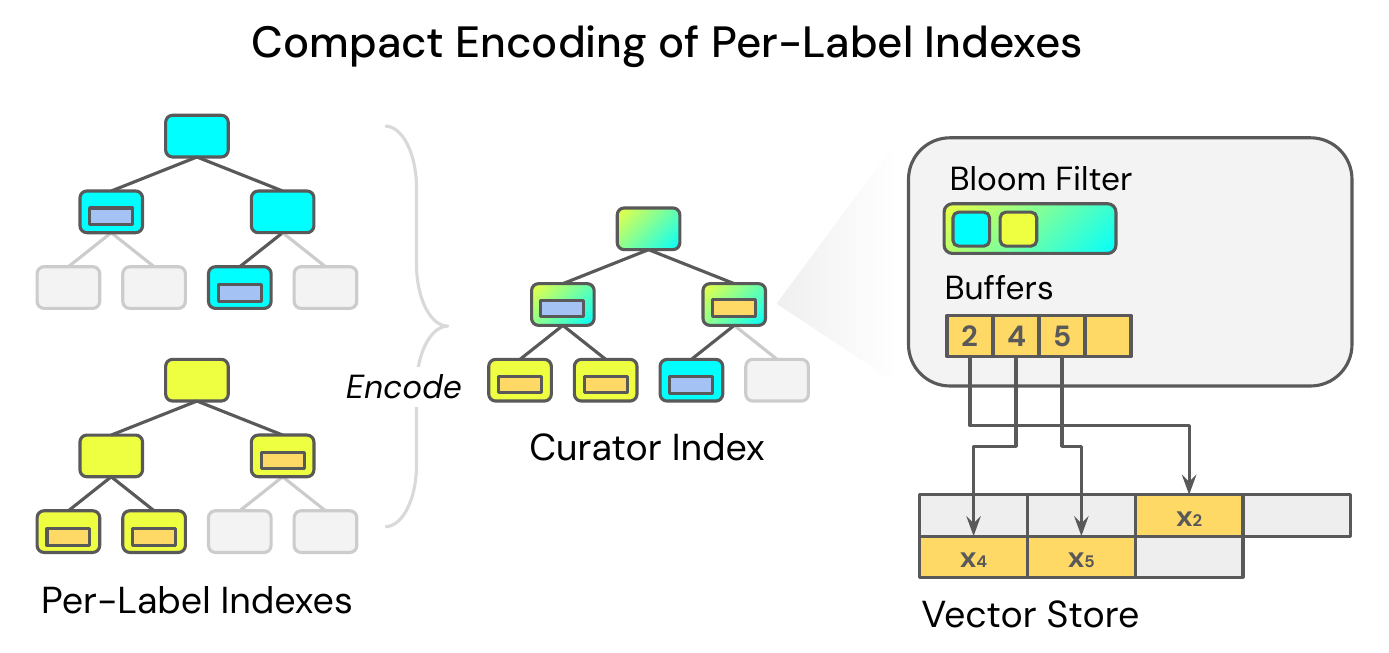}
  \caption{Overview of \sys. The base index represents the finest-grained partitioning on unfiltered vector distribution, while the per-label indexes are embedded within the base index, adapting to the unique distribution of their respective labels. Per-label indexes are compactly encoded using buffers and Bloom filters.}
  \label{fig:overview}
\end{figure}

This section presents \sys's architecture and design principles. Fig.~\ref{fig:overview} illustrates \sys's core design: a single shared \emph{base index} and a set of \emph{per-label indexes} embedded within it. This strategy allows per-label indexes to share the base index's spatial partitioning while adapting their granularity to their unique label distributions, addressing the fundamental limitations of existing approaches for low-selectivity queries outlined in \autoref{sec:background}.

\subsection{Base Index}

The base index $T$ is a standard \emph{hierarchical k-means clustering tree}~\cite{muja2014scalable} that partitions all vectors in the dataset. The tree recursively partitions the vector space using k-means clustering until reaching sufficiently fine granularity, i.e., when a node contains fewer vectors than the predefined \emph{leaf capacity}. The structure of the base index thus adapts to the distribution of all indexed vectors, which we refer to as the \emph{vector distribution}. Within the base index, each node stores a centroid that represents the corresponding cluster during search, while leaf nodes store the actual vector data within the cluster. This base index serves as the shared partitioning structure for embedded per-label indexes.

We choose hierarchical k-means as the base index for two key reasons. First, as a partition-based index, it does not suffer from the inherent connectivity issues that plague graph-based indexes at low selectivity, as partitions remain well-defined regardless of label sparsity. Second, the hierarchical structure naturally embeds varied partitioning granularities for different filters within a single tree, allowing per-label indexes to adapt their clustering depth to label-specific distributions. Regarding search performance, hierarchical k-means is typically slower than graph-based indexes like HNSW but comparable to IVF for unfiltered search. This performance positioning is confirmed in our evaluation (Fig.~\ref{fig:recall_vs_latency_arxiv}), where \sys traversing per-label clustering trees achieves performance comparable to per-label IVF.

\subsection{Per-Label Indexes}

Per-label indexes are embedded within the base index to adapt to the unique distribution of their respective labels. Unlike traditional approaches that build separate indexes for each label, \sys's per-label indexes reuse the base index's spatial partitioning while adjusting granularity to label-specific vector densities. This hierarchical storage strategy ensures that all vector data resides exclusively in the base index, avoiding the costly vector duplication that would occur if per-label indexes stored their own copies of vector data.

Per-label indexes are compactly encoded using two key data structures: \emph{buffers} that store vector identifiers at leaf nodes, and \emph{Bloom filters} that define the structure of per-label indexes.

\parab{Vector Identifiers and Buffers.} To avoid vector duplication, per-label indexes store buffers—sorted lists of \emph{vector identifiers}—at leaf nodes rather than actual vectors. The \emph{buffer capacity} $B_{max}$ controls leaf node granularity. These identifiers are constructed hierarchically to encode each vector's tree location: each node is assigned a \emph{node identifier} as a sequence of branch indices from the root, and each vector's identifier combines its containing leaf's node identifier with its index within that leaf. These two components are bit-packed into a single integer with zero padding for uniform length. This construction ensures each node identifier represents a \emph{vector ID subspace} containing all vectors with that prefix, so sorting vector identifiers groups vectors from the same sub-tree contiguously (Fig.~\ref{fig:temp_index}). This property enables efficient construction of temporary per-predicate indexes (\autoref{sec:algorithm:complex_predicate}), batch construction of per-label indexes (\autoref{sec:algorithm:rebuild}), and buffer split/merge operations during updates (\autoref{sec:algorithm:update}).

\parab{Bloom Filters.} The base index contains three distinct node types with respect to each per-label index: \emph{internal nodes} requiring further partitioning, \emph{leaf nodes} containing buffers with qualified vectors, and \emph{external nodes} not included in the per-label index. During search, \sys must efficiently determine the boundaries of each per-label index—consisting of leaf nodes and external nodes—to know when to backtrack. While leaf nodes are easily identified by the presence of buffers, distinguishing between internal and external nodes is challenging since both lack buffers. 

A straightforward approach would maintain an exact set at each node $n$ representing the per-label indexes containing it, denoted as $\mathcal{T}_n = \{l \mid n \in T_l\}$. During search, checking membership $l \in \mathcal{T}_n$ would determine whether the current node is inside $T_l$. However, storing $\mathcal{T}_n$ exactly becomes costly when the number of unique labels in the dataset is large. Therefore, \sys approximates $\mathcal{T}_n$ using Bloom filters $\mathcal{B}_n$ at each node in the base index. Bloom filters are space-efficient probabilistic data structures that support membership queries with bounded false positive rate $p$. Importantly, Bloom filters have no false negatives, ensuring that when $l \notin \mathcal{B}_{n}$, the search algorithm has definitively reached an external node and can safely backtrack. While false positives may cause occasional unnecessary exploration of nodes outside the per-label index, they do not affect search correctness since external nodes do not contain buffers and thus never contribute to the result set.

\begin{table}[t]
\centering
\caption{Notations used in the paper}
\label{tab:terms}
\small
\begin{tabular}{cll}
\toprule
& \textbf{Notation} & \textbf{Description} \\
\midrule
& \( \mathcal{P} \) & Set of indexed vectors \\
& \( \mathcal{L} \) & Finite universe of labels \\
& \( \mathcal{P}_l \) & Vectors associated with label \( l \) \\
& \( \mathcal{L}_x \) & Labels associated with vector \( x \) \\
& \( T, T_l \) & Index and per-label index for label \( l \) \\
& \( \mathcal{P}_n \) & Cluster represented by tree node \( n \) \\
& \( \mathbf{\mu}_n \) & Centroid of node \( n \) \\
& \( \mathcal{P}_{n,l} \) & Buffer for label \( l \) at node \( n \) \\
& \( \mathcal{B}_n \) & Bloom filter at node \( n \) \\
& \( C_n \) & Child nodes of node \( n \) \\
& \( B_{max} \) & Buffer capacity \\
\bottomrule
\end{tabular}
\end{table}

\section{Search Algorithms}
\label{sec:algorithms}

This section presents the core search algorithms that enable \sys to deliver efficient search when the filter selectivity is low. We distinguish two query types: single-label search and complex predicates that involve more attribute types and logical operators. The embedded per-label index structure ensures both \emph{efficiency} by avoiding visits to unqualified vectors and adapting structure to label distribution, and \emph{completeness} by covering all qualifying vectors within a connected clustering tree. For single-label filters, \sys traverses the corresponding pre-built per-label index; for complex predicates, it first constructs a temporary per-predicate index with minimal overhead and then applies the exact same search algorithm.

\subsection{Single-Label Search}
\label{sec:algorithm:simple_filter}

\begin{figure}[t]
  \centering
  \includegraphics[width=0.75\columnwidth]{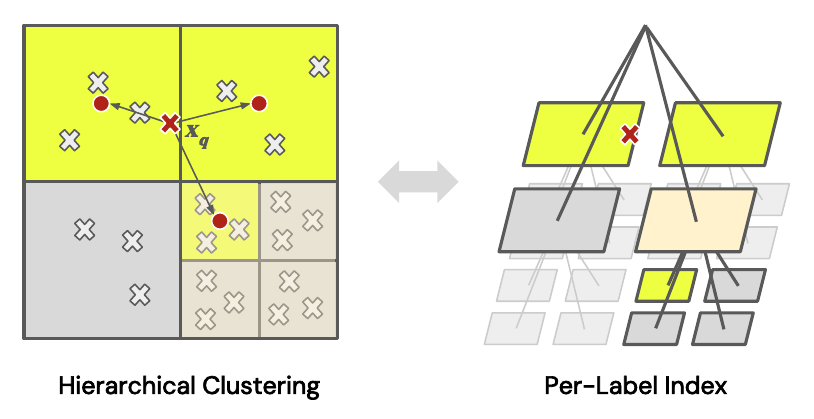}
  \caption{Single-label search on \sys. Highlighted clusters and tree nodes are visited by the search algorithm.}
  \label{fig:query}
\end{figure}

Single-label search finds qualifying vectors by scanning clusters closest to the query vector within the space partitioning represented by the per-label index (Fig.~\ref{fig:query}). Algorithm~\ref{algo:simple_filtered_search} implements this in two phases: beam search for robust initialization that traverses to the neighborhood of the query vector, followed by best-first search for efficient exploration of that neighborhood.

\begin{algorithm}[tb]

\DontPrintSemicolon

\SetKw{Break}{break}
\SetKw{Continue}{continue}
\SetKwFunction{BeamSearch}{BeamSearch}
\SetKwInOut{Input}{Input}\SetKwInOut{Output}{Output}

\caption{Single-Label Search}
\label{algo:simple_filtered_search}
\Input{Index $T$, query vector $x_q$, label $l$, beam width $b$, result set size $\mathit{ef}$}
\Output{$k$ approximate nearest neighbors of $x_q$}

Initialize $\mathcal{Q} \leftarrow \BeamSearch(T.\mathit{root}, b)$ \tcp*{Frontier}
Initialize $\mathcal{R} \leftarrow \varnothing$ \tcp*{Result set}

\While{$\mathcal{Q} \ne \varnothing$}{
    $n \leftarrow \arg \min _{n^\prime \in \mathcal{Q}} S(n^\prime, x_q)$\;
    $\mathcal{Q} \leftarrow \mathcal{Q} \setminus \{n\}$\;
    \uIf(\tcp*[f]{\circled{1} Outside $T_l$}){$l \notin \mathcal{B}_n$}{
        \Continue\;
    }
    \uElseIf(\tcp*[f]{\circled{2} Leaf node}){$\mathcal{P}_{n,l} \ne \varnothing$}{
        $\mathcal{R}^\prime \leftarrow \mathcal{R} \cup \mathcal{P}_{n,l}$\;
        \If{$|\mathcal{R}^\prime| > \mathit{ef}$}{
            $\mathcal{R}^\prime \leftarrow $ top-$\mathit{ef}$ closest vectors in $\mathcal{R}^\prime$\;
        }
        \If{$\mathcal{R}^\prime = \mathcal{R}$}{
            \Break\;
        }
        $\mathcal{R} \leftarrow \mathcal{R}^\prime$\;
    }
    \Else(\tcp*[f]{\circled{3} Internal node}){
        $\mathcal{Q} \leftarrow \mathcal{Q} \cup C_n$\;
    }
}
\Return{top-$k$ closest vectors in $\mathcal{R}$}
\end{algorithm}

\parab{Phase 1: Beam Search.}
Algorithm~\ref{algo:simple_filtered_search} begins with beam search to initialize the search frontier $\mathcal{Q}$ with tree nodes representing nearest clusters. This phase traverses from the root until reaching leaf nodes of the per-label index, maintaining the top-$b$ closest nodes at each level. Without beam search, best-first search starting from the root would be misguided by the imprecision of high-level nodes, causing the algorithm to become trapped in suboptimal subtrees without exploring subtrees whose centroids are relatively far from the query vector but still contain nearby vectors within their boundaries. Exploring multiple promising paths simultaneously mitigates this centroid imprecision by increasing the likelihood that the actual nearest leaf nodes are covered by the beam search.

\parab{Phase 2: Best-First Search.}
Using the frontier initialized by beam search, the second phase performs best-first search, systematically exploring the neighborhood guided by heuristic scores. The algorithm processes nodes based on their type with regard to the per-label index $T_l$: \emph{leaf nodes} containing buffers update the result set with their qualified vectors. For nodes without buffers, the Bloom filter $\mathcal{B}_n$ determines whether the node is an \emph{internal node} requiring expansion or an \emph{external node} outside $T_l$ where the search should backtrack. During exploration, the search maintains a dynamic result set $\mathcal{R}$ containing the $\mathit{ef}$ closest vectors discovered so far. Each time a buffer is visited, the algorithm updates $\mathcal{R}$ and checks for convergence—if the result set remains unchanged (i.e., no vector in the buffer is closer than the worst member of $\mathcal{R}$), the search terminates. Larger $\mathit{ef}$ values widen the exploration radius, trading more work for higher recall. In the limit, when $\mathit{ef}$ is sufficiently large, the traversal visits the entire per-label index encompassing all qualifying vectors, thus guaranteeing completeness. We found this adaptive termination criterion accommodates varying query difficulties more effectively than approaches that use a fixed computational budget.

\parab{Heuristic Function.}
Both search phases employ the same heuristic function to score candidate nodes: $S(n, x_q) = \left\|\mu_n-x_q\right\| - \alpha \cdot \frac{1}{|\mathcal{P}_n|} \sum _{x \in \mathcal{P}_n} \left\|x-\mu_n\right\|$. This heuristic estimates the minimum query-to-cluster distance by combining centroid distance with cluster radius. The $\alpha$ coefficient encourages exploring large, distant clusters to avoid local optima.

\subsection{Complex Predicate Search}
\label{sec:algorithm:complex_predicate}

The search algorithm described in the previous section supports only single-label categorical filters, which limits query expressiveness in two key ways: (1) it only supports categorical attributes, preventing range queries on numerical attributes, and (2) it does not support logical operators like AND, OR, or NOT to compose multiple filtering conditions.

\parab{Approach Overview.}
To address these limitations, we extend the system to support arbitrary complex predicates while maintaining the performance characteristics of single-label search. The key insight is that any complex predicate ultimately identifies a subset of vectors that qualify the given conditions. Instead of modifying the existing search infrastructure, we treat this subset as if it were associated with a virtual label, allowing us to reuse the same hierarchical partitioning and search algorithms. By constructing a temporary index containing only these qualifying vectors, we can apply the same search algorithm used for single-label filters.

The challenge lies in efficiently constructing this temporary index without expensive operations. A naive approach would require traversing the index for each qualified vector to find its corresponding position in the temporary index, which involves distance computations and tree traversals.

Our approach exploits a key property of the vector identifier construction described in \autoref{sec:overview}: given a list of qualified vectors sorted by their IDs, vectors from the same sub-tree are grouped contiguously. This property enables efficient temporary index construction through binary search operations. Given a sorted list of qualifying vector identifiers, we can recursively partition them into sub-trees by finding the boundaries between different child nodes using binary search, following the same hierarchical structure as the main index.

The sorted list of vector IDs, which we expect as the input for temporary index construction, is typically produced by an external relational database. This list could be represented as a posting list from an inverted index or a compressed bitmap. Evaluating attribute predicates and composing bitmaps inside the database is far cheaper than vector distance computation, so we treat this sorted list as the algorithm input. For further optimization, the costs of filter evaluation and temporary index construction can be reduced by only performing these operations in the sub-trees projected to be searched by the algorithm, rather than evaluating the predicate over all vectors and constructing the complete temporary index upfront.

\begin{algorithm}[tb]
\DontPrintSemicolon
\SetKw{Return}{return}

\caption{Temporary Index Construction}
\label{alg:temp_index}

\KwIn{Sorted qualified vector IDs $\mathcal{P}_{\sigma}$, main index $T$}
\KwOut{Temporary hierarchical index $T_{\sigma}$}

$T_{\sigma} \leftarrow$ \textsc{InitializeIndex}() \;
$T_{\sigma}.\mathit{root} \leftarrow$ \textsc{BuildSubtree}$(\mathcal{P}_{\sigma}, T.\mathit{root}, 0, |\mathcal{P}_{\sigma}|)$ \;
\Return{$T_{\sigma}$} \;

\vspace{0.5em}

\SetKwFunction{BuildSubtree}{BuildSubtree}
\SetKwProg{Fn}{Function}{}{}

\Fn{\BuildSubtree{$\mathcal{P}_{\sigma}$, $n$, $l$, $r$}}{
    \tcc{Build sub-tree rooted at $n$ for $\mathcal{P}_{\sigma}[l:r]$}
    $n_{\sigma} \leftarrow$ \textsc{InitializeNode}() \;

    \If(\tcp*[f]{\circled{1} Leaf node of $T_{\sigma}$}){$r - l \leq B_{max}$}{
        \Return{$n_{\sigma}$} \;
    }
    
    \ForEach(\tcp*[f]{\circled{2} Split $\mathcal{P}_{\sigma}[l:r]$}){$c \in C_n$}{
        \tcc{$c$ covers vector ID range $[c.\mathit{min}, c.\mathit{max})$}
        $l_c \leftarrow$ \textsc{BinarySearch}$(\mathcal{P}_{\sigma}, c.\mathit{min}, l, r)$ \;
        $r_c \leftarrow$ \textsc{BinarySearch}$(\mathcal{P}_{\sigma}, c.\mathit{max}, l, r)$ \;
        \If{$l_c \leq r_c$}{
            $c_{\sigma} \leftarrow$ \textsc{BuildSubtree}$(\mathcal{P}_{\sigma}, c, l_c, r_c)$ \;
            $n_{\sigma}.\textsc{AddChild}(c_{\sigma})$ \;
        }
    }
    
    \Return{$n_{\sigma}$} \;
}

\end{algorithm}

\begin{figure}[t]
  \centering
  \includegraphics[width=0.75\columnwidth]{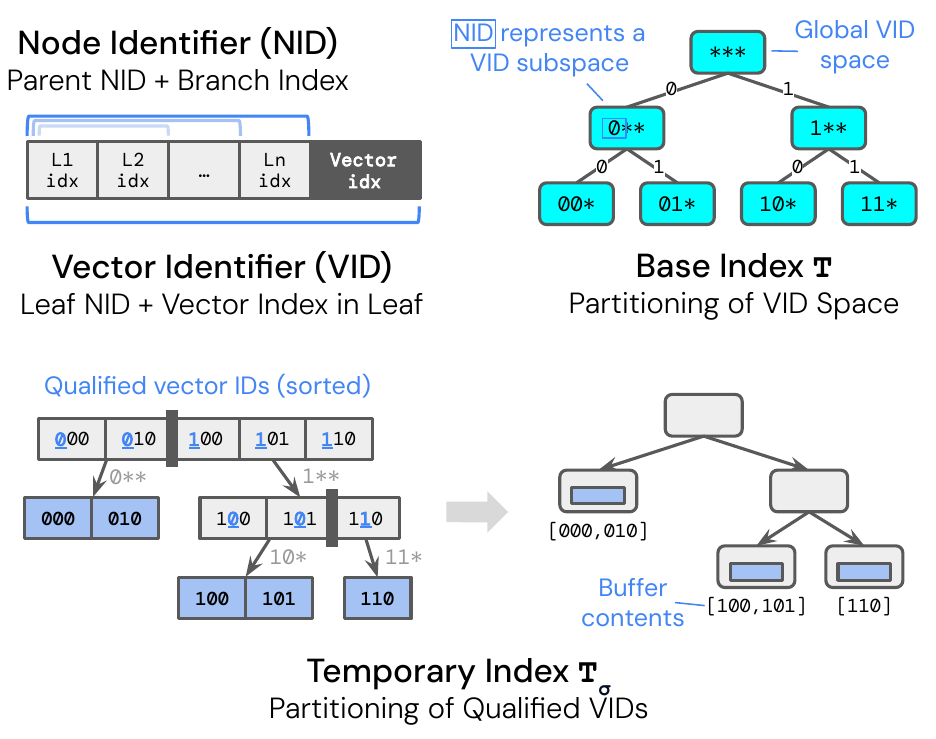}
  \caption{Temporary index construction for complex predicates. The sorted qualified vector IDs are recursively partitioned following the base index structure, enabling efficient construction of a temporary index.}
  \label{fig:temp_index}
\end{figure}

\parab{Temporary Index Construction.}
Algorithm~\ref{alg:temp_index} presents the temporary index construction procedure. The algorithm exploits the sorted vector ID property through a recursive tree building process that mirrors the structure of the main index $T$. The \textsc{BuildSubtree} function constructs each node $n_{\sigma}$ in the temporary index $T_{\sigma}$ by processing a corresponding range $\mathcal{P}_{\sigma}[l:r]$ of sorted qualified vector IDs. The algorithm operates as follows for each node: (1) Leaf condition: If the number of qualified vectors in the current range does not exceed the buffer capacity $B_{max}$, the algorithm creates a leaf node and terminates recursion for this sub-tree. (2) Hierarchical partitioning: Otherwise, it iterates through each child $c$ of the corresponding main index node $n$. For each child, binary search operations locate the subset of qualified vectors whose IDs fall within the child's vector ID range $[c.\mathit{min}, c.\mathit{max})$. This partitioning step leverages the contiguous property of sorted vector IDs to efficiently split the qualified vector list. After partitioning, the construction algorithm is then applied recursively to build the sub-tree for each child.

Fig.~\ref{fig:temp_index} illustrates this process using a simplified scenario where each node has two children and both leaf capacity and buffer capacity are 2, allowing vector IDs to be represented as 3-digit binary numbers. The left panel shows the base index with vector ID ranges annotated for each node, such as $[000\text{-}111]$ for the root and $[000\text{-}011]$, $[100\text{-}111]$ for its children, corresponding to the $[c.\mathit{min}, c.\mathit{max})$ ranges referenced in the algorithm. The middle panel demonstrates the recursive partitioning process: starting with sorted qualified vector IDs $\{000, 010, 100, 101, 110\}$, binary search finds the partition boundary at vector ID $100$ (the start of the right sub-tree's range $[100\text{-}111]$), splitting the list into groups $\{000, 010\}$ and $\{100, 101, 110\}$. The partitioning continues recursively until the leaf condition is met---for instance, the right sub-tree is further split into $\{100, 101\}$ and $\{110\}$, with the latter qualifying the leaf condition ($\leq 2$ vectors). The right panel shows the resulting temporary index where the qualified vector IDs are distributed into buffers in leaf nodes.

The temporary index maintains a lightweight structure where each node stores only: (1) a pointer to the corresponding node in the base index, (2) the range offsets $[l, r)$ in the qualified vector ID list, and (3) pointers to child nodes. This design minimizes memory overhead while preserving the hierarchical structure necessary for efficient search.

\parab{Search Process.}
Once the temporary index is constructed, the search algorithm proceeds identically to the single-label search case. While there are minor layout differences between the per-label indexes embedded in the main index and the constructed temporary index, they are conceptually equivalent, thus the difference does not affect the correctness or efficiency of the search algorithm.

\parab{Pre-indexing Filters.}
Despite finding that the overhead of temporary index construction is minimal in our evaluation, in certain scenarios it may be desirable to further eliminate this overhead by pre-constructing indexes for filter predicates, particularly when some filter predicates are frequently encountered. The simplest approach is caching the constructed temporary indexes as a side product of complex predicate search, which requires minimal additional implementation effort. Alternatively, we can permanently integrate the temporary indexes into the main index structure to save memory and reuse update routines. This integration process involves assigning a virtual label to the predicate to reference the temporary index in future queries, distributing the qualifying vectors as buffers across the appropriate nodes in the hierarchy, and updating the Bloom filters accordingly. The integration is straightforward due to the one-to-one mapping between nodes of the temporary index and the main index. Both approaches achieve nearly identical search performance in our evaluation, suggesting that the choice between them can be made based on application-specific considerations.

\sys treats pre-construction or caching of temporary indexes as a pluggable policy. Deployments may supply custom logic or rely on standard workload analysis tools and query optimizers to identify frequent, high-payoff predicates within resource budgets, preserving compatibility with existing index management pipelines.

\section{Index Maintenance}
\sys maintains both a base index fitted to the overall vector distribution and multiple per-label indexes adapted to their respective label distributions. As data evolves, the system must efficiently update these structures while preserving search performance. This section describes the incremental update mechanisms and rebuilding strategies that maintain index quality with minimal overhead.

\subsection{Incremental Updates}
\label{sec:algorithm:update}

\begin{figure}[t]
    \centering
    \includegraphics[width=0.85\columnwidth]{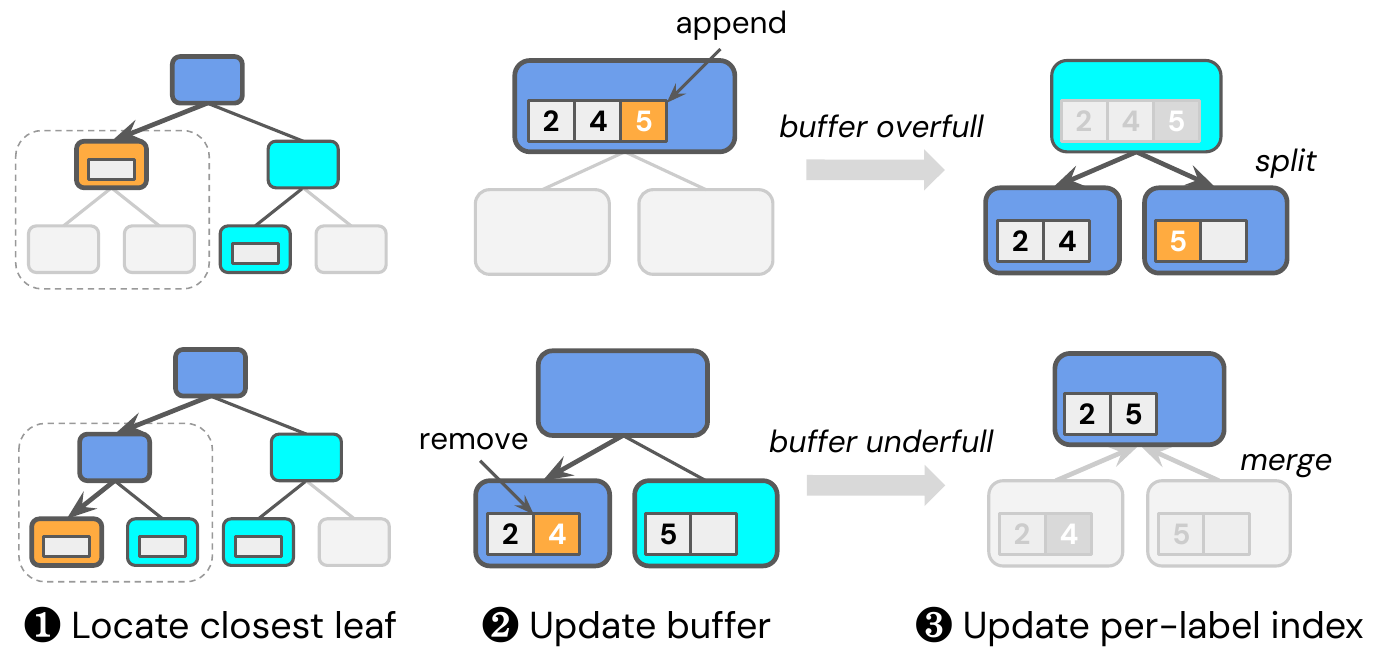}
    \caption{Label update operations in \sys. Top: label insertion triggers buffer overflow, causing vectors to be flushed to child nodes. Bottom: label deletion creates underfull buffers in child nodes that merge into the parent node.}
    \label{fig:label_update}
\end{figure}

\sys supports efficient incremental updates to both vectors and labels, each requiring different maintenance strategies.

\parab{Label Updates.}
Label updates modify per-label indexes through buffer operations and Bloom filter maintenance, as illustrated in Fig.~\ref{fig:label_update}. Throughout this process, \sys maintains two index invariants that ensure both search performance and correctness: buffer capacity limits and Bloom filter correctness.

\sys maintains buffer capacity limits using a strategy similar to buffer trees~\cite{arge1995buffer}. When a buffer overflows due to label insertion, vectors are flushed to child nodes at the next level (Fig.~\ref{fig:label_update} top). Conversely, when sibling buffers' combined size falls below capacity due to deletions, they merge into the parent node (Fig.~\ref{fig:label_update} bottom). These operations occur recursively and avoid distance computations by leveraging the root-to-leaf path encoded in vector identifiers. Efficiency is further enhanced by the contiguous grouping property (\autoref{sec:overview}), where vectors belonging to the same child are always grouped contiguously within buffers, enabling efficient partitioning via array slicing.

\sys maintains Bloom filter correctness when buffer operations cause structural changes to per-label indexes. For label insertions that expand the per-label index to new nodes, the label is simply added to the Bloom filter at those nodes. For buffer merging that removes nodes from per-label indexes, the Bloom filter of parent node $p$ must be recomputed recursively based on its children $C_p$, since Bloom filters cannot remove elements:
\begin{equation}
    \mathcal{B}_p = \{l \mid l \in \mathcal{L}, \mathcal{P}_{p,l} \neq \varnothing\} \cup \bigcup_{c \in C_p} \mathcal{B}_c
\end{equation}
Here, the first term represents labels with buffers stored at $p$ (per-label trees where $p$ is a leaf node), while the second aggregates child Bloom filters (per-label trees where $p$ is an internal node). Updates propagate recursively from modified nodes upward until no further changes occur.

\parab{Vector Updates.}
Unlike label updates, vector updates modify only the base index by finding the nearest leaf node through greedy search and adding or removing vector identifiers there. For deletions, the root-to-leaf path encoded in vector identifiers eliminates the need for greedy search, as the target leaf node can be directly identified from the vector's encoded path. Vector updates could use a similar split/merge strategy as label updates, but this would require updating multiple per-label indexes affected by the structural changes. This impact is typically minimal since few per-label indexes reach the leaf nodes of the base index due to label sparsity, but \sys instead enforces leaf capacity through periodic rebuilding (\autoref{sec:algorithm:rebuild}) for simplicity.

\subsection{Index Rebuilding}
\label{sec:algorithm:rebuild}
While incremental updates handle individual changes efficiently, accumulated updates can cause bucket imbalances that degrade search performance~\cite{xu2023spfresh}. To address this, index rebuilding involves reconstructing either the entire base index or a sub-tree, followed by batch construction of per-label indexes within the affected region.

\sys offers two rebuilding strategies that provide different trade-offs between overhead and freshness. \textit{Global rebuilding} reconstructs the entire base index using hierarchical k-means clustering, followed by batch construction of all per-label indexes. The batch construction of per-label indexes follows the same recursive partitioning procedure as temporary index construction for complex predicates (\autoref{sec:algorithm:complex_predicate}). This provides optimal index quality but incurs higher overhead to maintain the same level of freshness. In contrast, \textit{local rebuilding} targets only sub-trees with significant updates, reducing overhead while maintaining freshness where needed. To determine when local rebuilding is necessary, \sys tracks update statistics at each node: total vectors $|\mathcal{P}_n|$ and updates since last rebuild $U_n$. When the update ratio $U_n / |\mathcal{P}_n|$ exceeds a predefined threshold, the sub-tree enters a rebuilding queue for batch processing at defined intervals or on manual request, with new sub-trees atomically swapped in and counters reset.

\section{Evaluation}

This section empirically evaluates our method across various dimensions. Overall, our results demonstrate:

\begin{itemize}
    \item \sys achieves competitive search latency with minimal build time and index overhead compared to other approaches.
    
    \item \sys can be efficiently integrated into existing graph-based indexes with minimal overhead to address their performance issues at low selectivity. Supplementing ACORN with \sys improves search performance by up to 20.9$\times$ with merely 5.5\% and 4.3\% overhead in construction time and memory footprint, respectively.

    \item \sys efficiently supports incremental updates of both vectors and labels, as well as low-selectivity queries with arbitrary complex predicates.
\end{itemize}

\subsection{Experimental Setup}

\parab{Environment.} 
\sys and the baselines are implemented in C++ and compiled using GCC 9.4.0 with \texttt{-O3} and \texttt{AVX512} optimization. Our implementation builds upon the FAISS library~\cite{johnson2019faiss} infrastructure. All evaluations run on a server equipped with an Intel Xeon Gold 5215 @2.5GHz processor and 256GB of RAM, running Linux Ubuntu 20.04 LTS. Unless otherwise specified, all experiments are conducted in a single-threaded environment.

\parab{Datasets.}
We use the following real-world datasets for evaluation and provide their statistics in Table~\ref{tab:dataset_characteristics}. Each dataset is randomly divided into training and test sets, with the test set containing 10,000 vectors.
(1) The YFCC-10M dataset is a 10M vector subset of the original YFCC100M dataset~\cite{thomee2016yfcc100m}, which serves as the standard benchmark in the NeurIPS'23 Big-ANN competition~\cite{NeurIPS2023BigANN}. This dataset features a hybrid of vector and scalar data, with images embedded using the CLIP model~\cite{radford2021learning} and annotated with labels representing objects, locations, etc. We use the 1000 most popular labels to ensure all baselines consume a manageable amount of memory.
(2) The arXiv dataset~\cite{clement2019usearxiv} contains metadata from 2.3M arXiv papers, including titles, authors, abstracts, and paper categories (e.g., \verb|cs.DB| denotes the field of database systems). We embed paper abstracts using the \verb|all-MiniLM-L6-v2|~\cite{huggingface_allMiniLM-L6-v2} model and use the 100 most popular categories as labels.

\begin{table}[t]
\centering
\caption{Characteristics of Evaluation Datasets}
\label{tab:dataset_characteristics}
\begin{tabular}{lcc}
\toprule
                   & \textbf{YFCC-10M}  & \textbf{arXiv} \\
\midrule
\textbf{Source Data}& Image & Text      \\
\textbf{\# Vectors}& 10M      & 2M        \\
\textbf{Vector Dimension}& 192     & 384       \\
\textbf{Unique Labels}& 1000    & 100       \\
\textbf{Labels per Vector}& 5.53   & 9.93       \\
\bottomrule
\end{tabular}
\end{table}

\parab{Baselines.}
We evaluate the performance of \sys against the following indexes: IVF~\cite{johnson2019faiss}, HNSW~\cite{malkov2018hnsw}, Parlay-IVF~\cite{parlayivf}, Filtered DiskANN~\cite{gollapudi2023filtered}, and ACORN~\cite{patel2024acorn}. IVF and HNSW represent state-of-the-art partition-based and graph-based indexes, respectively. For both IVF and HNSW, we develop variants of per-label indexing and inline filtering to facilitate filtered search. These variants are referred to as Per-label IVF/HNSW and Shared IVF/HNSW, abbreviated as P-IVF/HNSW and S-IVF/HNSW, respectively, throughout the paper. Parlay-IVF is the top open source entry in the filtered search track of the Big-ANN competition, which constructs per-label graph-based indexes with shared vector storage. 
Due to excessive memory requirements, we omit evaluation of Per-label IVF and Per-label HNSW baselines on the YFCC-10M dataset. We also exclude NHQ~\cite{wang2022navigable} from our evaluation as it requires specifying the presence or absence of all labels in filter predicates, and UNG~\cite{cai2024navigating} due to its poor performance on our datasets (as discussed in \autoref{sec:background}).

In addition to these standalone vector indexes, we evaluate \texttt{pgvector}~\cite{pgvector}, a PostgreSQL extension that allows users to create vector-typed columns and build vector indexes on them, integrating vector search with SQL-based predicate filtering. We evaluate two \texttt{pgvector}-based baselines: Pg-IVF and Pg-HNSW, which use IVF-Flat and HNSW as the underlying vector indexes, respectively. Vectors and labels are stored in PostgreSQL tables with labels represented as \texttt{INT[]} arrays indexed using GIN for efficient containment queries. We use \texttt{UNLOGGED} tables to reduce index construction time. For Pg-IVF, we scan all vectors in the $nprobe$ buckets closest to the query vector and return the top-$k$ qualified vectors. Distances are computed for both qualified and unqualified vectors. Pg-HNSW employs iterative search that resumes exploration from the next $\mathit{ef}$ best candidates after each search iteration converges, terminating after finding $k$ qualified vectors or scanning at most $T_{max}$ vectors. To generate recall-throughput curves, we sweep $nprobe$ for Pg-IVF and $T_{max}$ for Pg-HNSW.

In all experiments, we store vectors and compute distances in \verb|float32| precision without scalar or product quantization~\cite{jegou2010product}, and conduct parameter sweeps to determine the Pareto-optimal configurations for each index. For experiments displaying the performance of only a single configuration per index, we choose the configuration that achieves 90\% recall with the lowest search latency. We list the parameter space of each baseline below: (1)~Per-label IVF: $nlist \in \{10, 20, 40\}$, $nprobe \in \{1, 2, 4, 8\}$; (2)~Per-label HNSW: $efc \in \{16, 32, 64, 128\}$, $M \in \{8, 16, 32\}$, $ef \in \{8, 16,\allowbreak 32, 64, 128\}$; (3)~Shared IVF: $nlist \in \{1000, 2000,\allowbreak 4000, 8000, 16000\}$, $nprobe \in \{8, 16, 32, 64, 128\}$; (4)~Shared HNSW: $efc \in \{16, 32, 64, 128\}$, $M \in \{16, 32, 64\}$, $ef \in \{8, 16,\allowbreak 32, 64, 128\}$; (5)~Parlay-IVF: $M \in \{8, 16, 32\}$, $ef \in \{8,\allowbreak 16, 32, 64, 128\}$; (6)~Filtered DiskANN: $efc \in \{64, 128,\allowbreak 256, 512\}$, $M \in \{64, 128, 256, 512\}$, $ef \in \{64, 128,\allowbreak 256, 512, 1024\}$, $\alpha = 1.2$; (7)~ACORN-$\gamma$: $M \in \{16, 32, 64\}$, $\gamma \in \{10, 20, 40\}$, $M_{\beta} \in \{M, 2M, 4M\}$, $ef \in \{8, 16, 32,\allowbreak 64, 128\}$; (8)~ACORN-1: $M \in \{16, 32, 64\}$, $\gamma = 1$, $M_{\beta} = 2M$, $ef \in \{8, 16, 32,\allowbreak 64, 128\}$; (9)~\sys: $nlist \in \{16, 32\}$, $B_{max} \in \{64, 128, 256\}$ (buffer and leaf capacity), $ef \in \{64, 128, 256,\allowbreak 512, 1024\}$, $b = 4$ (beam size); (10)~Pg-IVF: $nlist \in \{1000, 2000,\allowbreak 4000, 8000, 16000\}$, $nprobe \in \{8, 16,\allowbreak 32, 64, 128,\allowbreak 256, 512\}$; (11)~Pg-HNSW: $efc \in \{16, 32, 64, 128\}$, $M \in \{16, 32, 64\}$, $ef = 128$, $T_{max} \in \{5 \times 10^3,\allowbreak 1 \times 10^4,\allowbreak 2 \times 10^4, 4 \times 10^4, 8 \times 10^4\}$.

\subsection{Query Performance}
\label{sec:eval:query_performance}

\begin{figure*}[t]
    \centering
    \includegraphics[width=\textwidth]{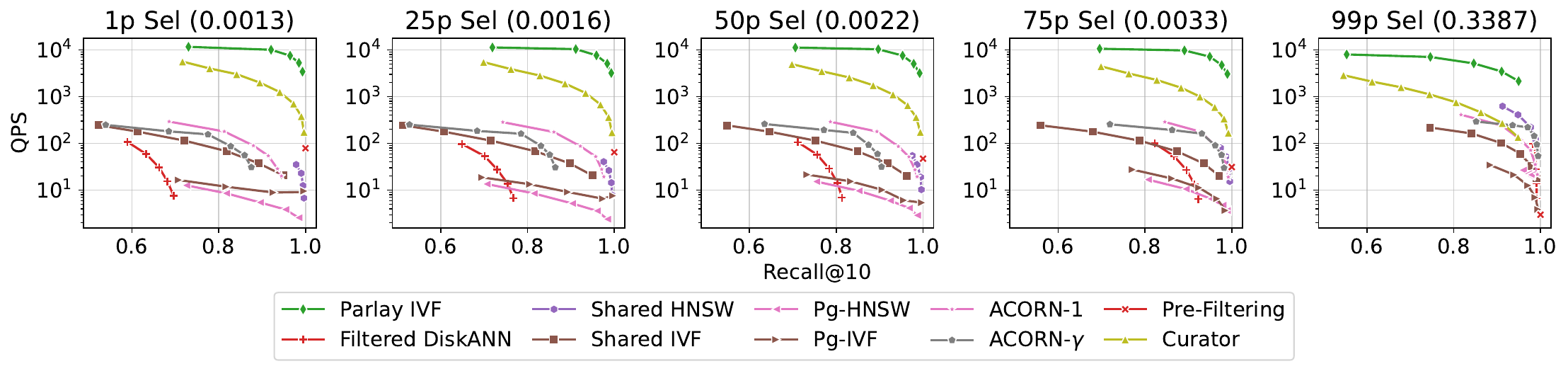}
    \caption{Trade-offs between recall and query throughput for single-label filters with varied selectivity on YFCC-10M. Each line represents an algorithm's performance across various search parameter configurations. Parenthetical numbers in subplot titles indicate the actual selectivity value for each filter.}
    \label{fig:recall_vs_latency_yfcc}
\end{figure*}

\begin{figure*}[t]
    \centering
    \includegraphics[width=\textwidth]{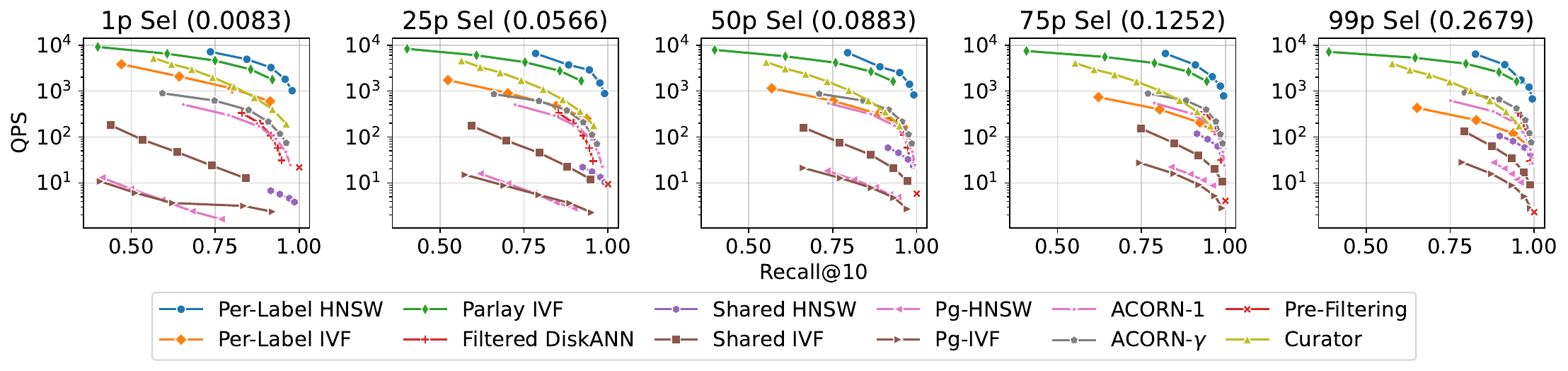}
    \caption{Trade-offs between recall and query throughput for single-label filters with varied selectivity on arXiv. Each line represents an algorithm's performance across various search parameter configurations. Parenthetical numbers in subplot titles indicate the actual selectivity value for each filter.}
    \label{fig:recall_vs_latency_arxiv}
\end{figure*}

\parab{Single-Label Search.}
\label{sec:eval:query_perf:simple_filter}
Fig.~\ref{fig:recall_vs_latency_yfcc}-\ref{fig:recall_vs_latency_arxiv} show the recall-throughput trade-offs across varied filter selectivity levels. Overall, per-label indexing approaches achieve the best search performance when feasible: Parlay-IVF consistently delivers optimal results by sharing vector data among per-label indexes, while Per-Label HNSW and IVF's performance on arXiv do not scale to the 10M-scale YFCC-10M dataset due to excessive memory overhead. However, these approaches come with prohibitive memory and construction costs, as demonstrated in our analysis in the subsequent sections, making them impractical for many real-world deployments.

For specialized indexes, performance degrades significantly as selectivity decreases due to graph connectivity issues. Filtered DiskANN's performance drops most severely, suffering from its less efficient compression strategy, while ACORN-1 slightly outperforms ACORN-$\gamma$ at very low selectivity levels. Metadata filtering baselines generally exhibit lower query throughput but show interesting behavior at high selectivity levels: Shared HNSW achieves better performance than specialized indexes at the 99th percentile selectivity (0.34) on YFCC-10M, demonstrating that the query-time overhead of specialized indexes can offset the benefits of fewer unnecessary distance computations. The \texttt{pgvector} baselines generally achieve the lowest query throughput across all evaluated approaches (except Filtered DiskANN on YFCC-10M at high recall regimes), which can be attributed to executor overhead and disk I/O costs that in-memory indexes do not incur. Notably, Pg-IVF exhibits an unusual pattern in low selectivity query groups: at high search budgets, the recall-throughput curve flattens. This occurs because the query planner estimates high cost for vector index scans and opts for pre-filtering instead.

\sys consistently achieves competitive performance across all selectivity levels, particularly excelling at low selectivity where specialized indexes and metadata filtering baselines struggle. For example, Shared HNSW and \texttt{pgvector} are dominated by pre-filtering at low selectivity, while \sys achieves higher throughput than pre-filtering at near-perfect recall and offers a flexible recall-throughput trade-off. Additionally, \sys matches Per-Label IVF's search performance on arXiv while requiring significantly lower memory usage and construction costs, demonstrating the efficiency of its embedded per-label indexes.

This performance profile reflects \sys's deliberate design trade-offs as discussed in \autoref{sec:intro}. Per-label indexing approaches (Parlay-IVF, Per-Label HNSW/IVF) achieve the best search performance by eliminating query-time filtering but require excessive memory overhead and construction costs due to vector duplication, as demonstrated in \autoref{sec:eval:efficiency}. Similarly, \sys exhibits slower search than specialized graph-based indexes (ACORN) at high selectivity, which is expected since \sys's partition-based approach prioritizes robustness to sparsity over the precision of graph connectivity. However, \sys's performance advantage emerges precisely where it was designed to excel: at low selectivity where graph connectivity breaks down, \sys maintains efficient search while other approaches either fail entirely or fall back to expensive pre-filtering.

\begin{figure}[t]
  \centering
  \includegraphics[width=\linewidth]{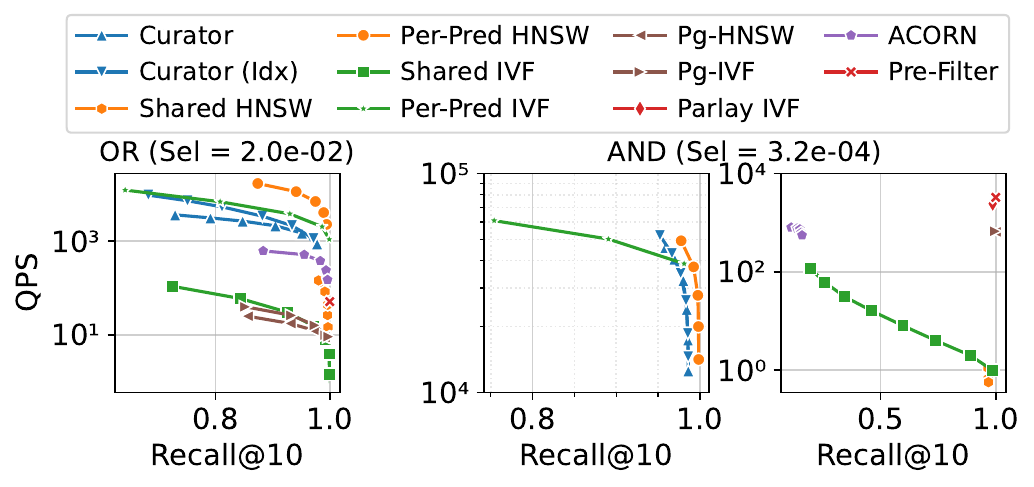}
  \caption{Recall vs query throughput trade-offs for complex predicate queries. Each line represents an algorithm's performance across various search parameter configurations. Average selectivity values are annotated in subplot titles. AND query results are split into two subplots due to the wide throughput range. Parlay-IVF only supports AND queries.}
  \label{fig:complex_predicate}
\end{figure}

\parab{Complex Predicate Queries.}
\label{sec:eval:query_perf:complex_predicate}
We construct queries with complex predicates based on the YFCC-10M dataset, evaluating two types of filter predicates: OR queries ($l_1 \vee l_2$, requiring existence of either label in the vector's label list) and AND queries ($l_1 \wedge l_2$, requiring existence of both labels). Given the average label selectivity of approximately 1\% in YFCC-10M, OR queries have selectivity around 2\% while AND queries have selectivity around 0.01\%. For each predicate type, we randomly generate 100 label combinations and select 100 query vectors from the test set, resulting in 10,000 total queries.

We evaluate \sys's performance against Shared HNSW, Shared IVF, Parlay-IVF (AND queries only), ACORN, and \texttt{pgvector} baselines (Pg-IVF and Pg-HNSW), which are the only baselines supporting complex predicates. Two variants of \sys are evaluated: (1) predicate not indexed, where \sys constructs a temporary index structure at query time for qualified vectors and traverses it, and (2) predicate indexed, where qualified vectors for each predicate are pre-indexed as virtual labels, eliminating the need for temporary index construction during search. Per-predicate HNSW and IVF approaches, which build separate indexes for all possible predicates, serve as references for optimal performance but are generally impractical due to prohibitive memory and construction costs. 

As shown in Fig.~\ref{fig:complex_predicate}, \sys with predicate indexed achieves near-optimal performance compared to the per-predicate approaches, demonstrating the index quality of \sys's temporary indexes. Furthermore, \sys without indexing achieves similar performance to the indexed version, significantly outperforming ACORN on both filter types and showcasing the efficiency of temporary index construction at query time. ACORN fails to achieve >0.2 recall on AND queries due to limited connectivity in the graph structure at low selectivity. For AND queries, \texttt{pgvector} shows only a single point as the PostgreSQL query planner picks pre-filtering plan due to the extremely low selectivity.

The results demonstrate that \sys provides a unified solution for both single-label and complex predicate search: while it delivers competitive performance on single-label filters across all selectivity levels, it uniquely excels at complex predicates where most existing approaches either fail entirely or require prohibitive memory overhead for per-predicate indexing.

\subsection{Resource Efficiency}
\label{sec:eval:efficiency}

\begin{figure}[t]
  \centering
  \includegraphics[width=\linewidth]{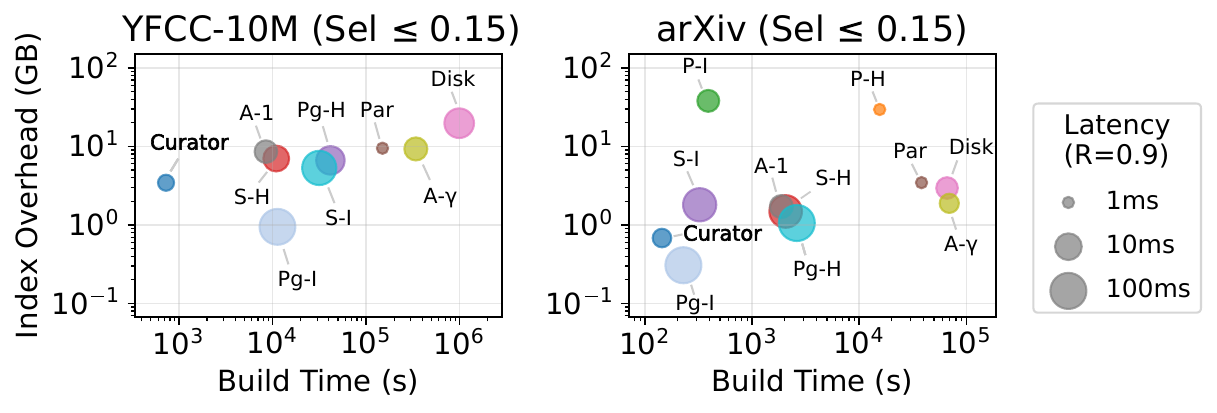}
  \caption{Trade-off between index overhead, build time and search latency for evaluated indexes. We report search latency at 0.9 recall on low selectivity queries ($\leq$ 0.15). For \texttt{pgvector} baselines, we report on-disk footprint rather than in-memory overhead. Marker size indicates search latency.}
  \label{fig:memory_vs_latency}
\end{figure}

\parab{Performance vs Resource Trade-offs.}
Fig.~\ref{fig:memory_vs_latency} provides a comprehensive view of the fundamental trade-offs between search performance, memory efficiency, and construction cost across all evaluated approaches. This analysis focuses on low-selectivity scenarios ($\leq$ 0.15) where the differences between approaches are most pronounced.

The results demonstrate that \sys achieves competitive search performance on low-selectivity queries while maintaining the lowest in-memory overhead and build time among all evaluated approaches. Only two baselines, Parlay-IVF and Per-Label HNSW, achieve lower search latency than \sys, but at significantly higher resource costs. Specifically, Parlay-IVF requires 206$\times$ and 266$\times$ longer build times on YFCC-10M and arXiv respectively, with 2.8$\times$ and 5.1$\times $ higher memory overhead. Per-Label HNSW (on arXiv) incurs 108$\times$ longer build time and 43$\times$ higher memory overhead for only marginal latency improvements. Pg-IVF's on-disk footprint is exceptionally low at 0.95GB on YFCC-10M and 0.31GB on arXiv, attributed to compact inverted list encoding. However, its search performance lags behind in-memory approaches due to executor overhead and disk I/O costs as discussed previously.

This analysis reveals why \sys represents an attractive practical solution: it delivers performance competitive with the best approaches while requiring minimal additional resources, making it suitable for deployment scenarios where memory and construction time constraints are critical considerations.

\begin{figure}[t]
  \centering
  \includegraphics[width=\linewidth]{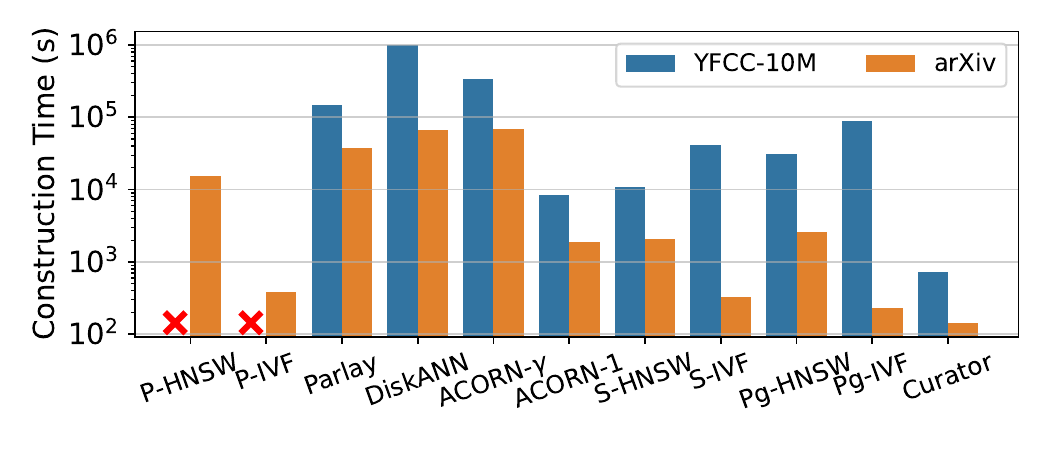}
  \caption{Index construction time for the evaluated indexes.}
  \label{fig:build_time}
\end{figure}

\parab{Construction Time.}
Fig.~\ref{fig:build_time} compares the index construction time of \sys against all baselines. The index construction time includes the time to optionally train the index structure (required for \sys and IVF baselines) and to insert all vectors and labels. Filtered DiskANN, ACORN, Parlay-IVF, and \texttt{pgvector} baselines are constructed in batch mode, whereas all other indexes are built incrementally.

Overall, \sys achieves the lowest construction time across all evaluated approaches, requiring only 727s on YFCC-10M and 144s on arXiv. Specialized indexes like ACORN-$\gamma$ and Filtered DiskANN incur dramatically higher construction costs on both datasets, primarily due to the overhead of constructing and compressing the intermediate dense graph. Specifically, ACORN-$\gamma$ requires 468$\times$ and 482$\times$ longer construction time than \sys on YFCC-10M and arXiv respectively, while Filtered DiskANN requires 1,360$\times$ and 459$\times$ longer time. These specialized approaches require significantly more construction time than even building separate per-label indexes: both ACORN-$\gamma$ and Filtered DiskANN take around 4$\times$ longer time to build than Per-Label HNSW on arXiv, highlighting the substantial overhead of these sophisticated indexing strategies.

\begin{figure}[t]
  \centering
  \includegraphics[width=\linewidth]{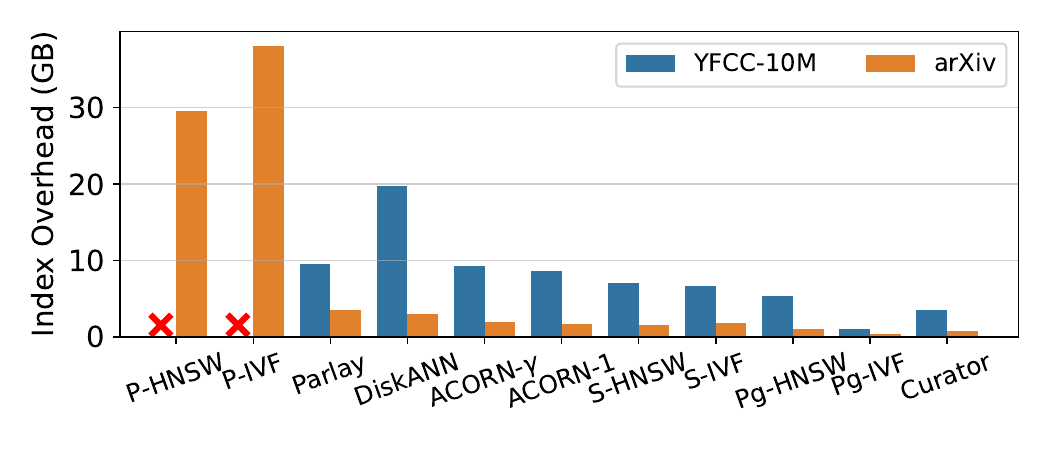}
  \caption{Index overhead for evaluated indexes. Index overhead is computed by subtracting the raw vector storage size (excluding label lists) from the total memory footprint. For \texttt{pgvector} baselines, we report on-disk footprint.}
  \label{fig:memory_footprint}
\end{figure}

\parab{Memory Overhead.}
Index overhead represents the additional memory required by indexes beyond storing the original vectors (assuming no compression), calculated by subtracting raw vector storage size from the total memory footprint. We subtract vector storage because all indexes maintain at least one copy of vector data, making this a fair comparison baseline. However, we do not subtract label/metadata storage since indexes preserve metadata in vastly different formats: some store plain label lists or inverted indexes, while per-label indexing baselines require no metadata storage at all.

As shown in Fig.~\ref{fig:memory_footprint}, \sys and Pg-IVF achieve the lowest index overhead among all evaluated approaches. \sys requires 3.45GB on YFCC-10M and 0.68GB on arXiv, while Pg-IVF achieves even lower on-disk footprint at 0.95GB and 0.31GB respectively. Per-label indexing approaches incur dramatically higher memory costs due to vector data duplication, with Per-Label HNSW and Per-Label IVF requiring 43$\times$ and 56$\times$ more overhead than \sys on arXiv, respectively. Specialized indexes show more moderate but still significant overhead increases due to their large graph degrees: Filtered DiskANN requires 5.7$\times$ and 4.4$\times$ more memory than \sys, while ACORN variants require 2.5--2.8$\times$ more overhead across both datasets.

\subsection{Integration with Graph-Based Indexes}
\label{sec:eval:sys_perf}

In this section, we evaluate how integrating \sys with graph-based indexes improves search performance at low selectivity without significantly increasing construction costs.

\parab{Dataset.}
For our evaluation, we construct a semi-synthetic dataset from a randomly sampled 1M subset of the YFCC-10M dataset to achieve systematic coverage of the low-selectivity regime with sufficient data points at each selectivity level for reliable evaluation. Specifically, we define 20 selectivity levels distributed on a logarithmic scale within the range $[0.001, 0.2]$. The logarithmic distribution ensures dense sampling at the lower end of the selectivity spectrum where \sys is designed to excel. For each selectivity level, we generate 10 labels by randomly sampling the corresponding fraction of vectors from the dataset. We determine the optimal build configuration for each index through grid search and evaluate queries under varying search configurations. For each selectivity level, we report the minimum query latency required to achieve an average recall of 90\%.

\parab{Baselines and Integration Strategy.}
We select ACORN as the graph-based index as it achieves best performance on our datasets. To illustrate the trade-offs of densifying graphs, we evaluate two ACORN-$\gamma$ constructions with $\gamma = 10$ and $\gamma = 40$, denoted as ACORN-10 and ACORN-40, respectively. For low-selectivity queries, we combine ACORN with either pre-filtering or \sys, resulting in four hybrid indexes in total. In our integration experiments, we assume users perform selectivity estimation and choose indexes based on fixed selectivity thresholds determined through offline profiling, similar to ACORN's approach. In practice, this selection logic can also be handled by external systems such as database query optimizers, allowing flexible integration strategies tailored to specific workload patterns.

\parab{Results.}
As shown in Fig.~\ref{fig:sys_perf}, densifying the graph-based index yields negligible improvements at low selectivity while significantly raising construction costs, increasing construction time by approximately 5$\times$ in our experiments. Integrating ACORN-10 with \sys increases construction time and memory footprint by only 5.5\% and 4.3\%, respectively, while improving search performance by up to 20.9$\times$ compared to ACORN with pre-filtering at low selectivity. These results demonstrate that \sys provides an efficient complementary solution for low-selectivity queries without imposing significant resource overhead.

\subsection{Update Performance}
\label{sec:eval:updates}

\begin{figure}[t]
  \centering
  \includegraphics[width=\linewidth]{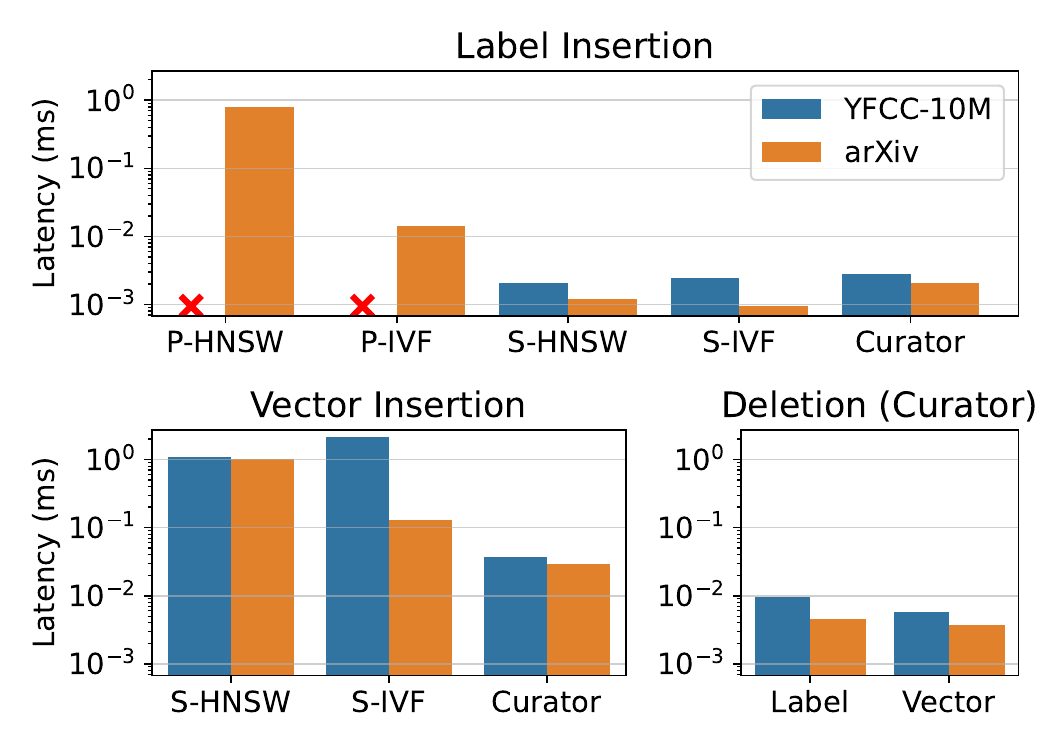}
  \caption{Update performance for the indexes supporting incremental updates. Per-label indexing baselines do not support vector insertion. Deletion latency for vectors and labels is only shown for \sys, as other indexes do not support deletion.}
  \label{fig:insert_performance}
\end{figure}

Fig.~\ref{fig:insert_performance} compares update performance of \sys against baselines supporting incremental updates. Parlay-IVF, Filtered DiskANN, and ACORN are excluded due to their lack of incremental update support. Per-label indexing baselines do not support vector insertion, as vectors are directly inserted into corresponding per-label indexes during label insertion operations. For deletion operations, only \sys supports both vector and label deletion, so performance is shown for \sys only.

While per-label indexing baselines avoid vector insertion costs, they incur significant overhead when assigning new labels to vectors, as each assignment involves an update to the corresponding index. In contrast, metadata filtering baselines involve costly distance computations during vector insertion but achieve fast label insertion by simply appending labels to metadata. Therefore, for scenarios where each vector is associated with many labels, the total update costs for per-label indexing are substantially higher than those for metadata filtering. 

Compared to all baselines, \sys achieves consistently low latency across all update operations. Vector insertion in \sys is 4.4--58.7$\times$ faster than shared index approaches, while label insertion maintains minimal overhead of 0.002--0.003 ms across both datasets. \sys's deletion performance is also efficient, with vector deletion requiring 0.004--0.006 ms and label deletion requiring 0.005--0.010 ms. This low latency stems from the efficient navigation afforded by \sys's tree structure and fast buffer operations.

\subsection{Scalability}
\label{sec:eval:scalability}

\begin{figure}[t]
  \centering
  \includegraphics[width=\linewidth]{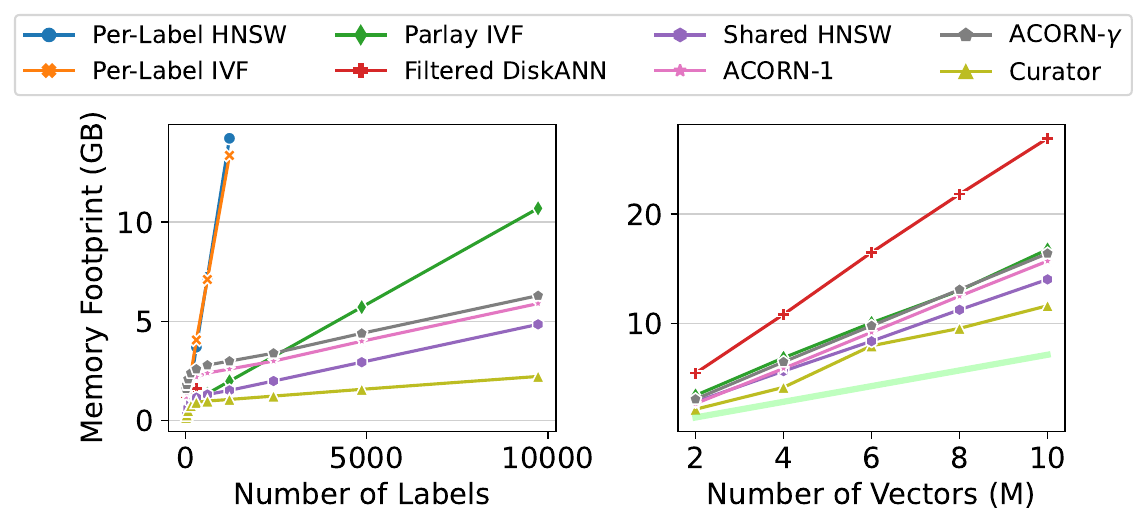}
  \caption{\sys exhibits superior memory efficiency as the number of labels and vectors scales. The light green line shows the size of vector storage alone.}
  \label{fig:scalability}
\end{figure}

In this section, we assess \sys's memory efficiency across varying numbers of labels and vectors.

\parab{Scaling with Number of Labels.} For this experiment, we use a 1M subset of YFCC-10M, generating a series of datasets with varying numbers of labels where each label is randomly assigned to 1\% of vectors. As shown in Fig.~\ref{fig:scalability} (left), per-label approaches (HNSW and IVF) exhibit the steepest growth due to vector duplication across multiple indexes. Parlay-IVF grows second fastest—while it shares vector storage, each vector is indexed multiple times as the label count increases, increasing the total number of graph edges. Shared HNSW and ACORN show moderate growth since their index structures remain unchanged as labels increase, requiring only additional label storage. \sys achieves the smallest growth rate through compact encoding of embedded per-label indexes. Filtered DiskANN fails beyond ~600 labels because the merged graph becomes increasingly dense, requiring more aggressive pruning to maintain a fixed maximum degree, which significantly degrades recall.

\parab{Scaling with Number of Vectors.} For this experiment, we use sampled subsets of the YFCC-10M dataset ranging from 2M to 10M vectors, with average labels per vector held constant. Per-label HNSW and IVF are excluded from this comparison due to excessive memory requirements. As shown in Fig.~\ref{fig:scalability} (right), the differences between graph-based approaches primarily stem from varied graph degrees: Filtered DiskANN requires significantly more edges to maintain connectivity after pruning, while Parlay-IVF, ACORN, and Shared HNSW exhibit similar memory growth rates since each vector maintains comparable total edge counts (aggregated across all per-label indexes for Parlay-IVF). \sys maintains the lowest memory footprint due to its compact encoding and partition-based design.

\subsection{Ablation Study}
\label{sec:eval:ablation}

\begin{figure}[t]
  \centering
  \includegraphics[width=\linewidth]{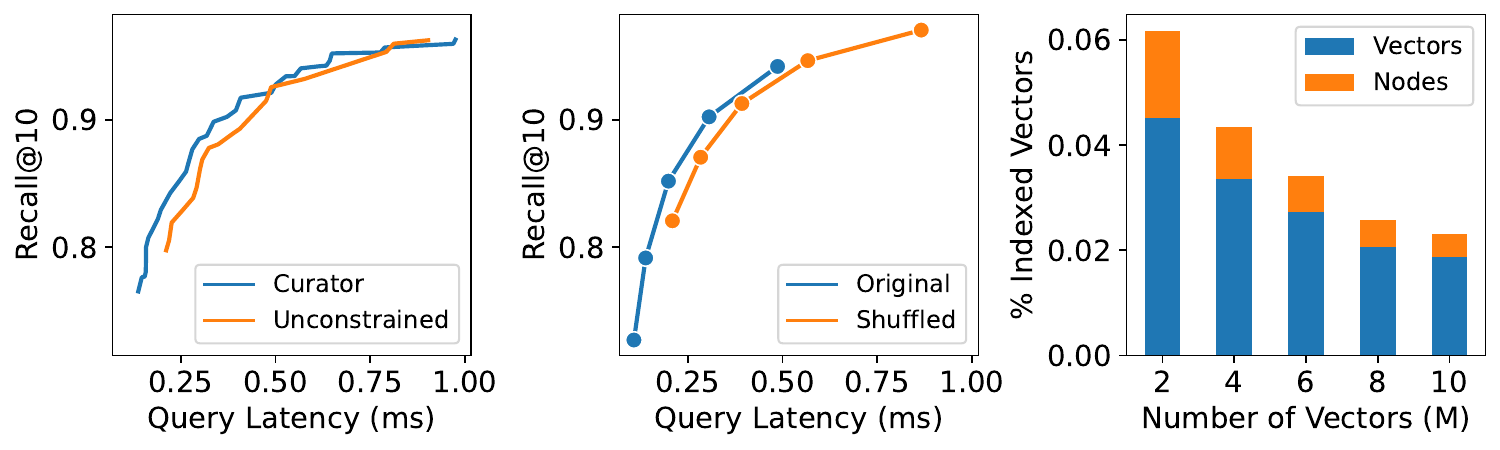}
  \caption{Ablation studies validating \sys's design: (a) Performance impact of structural constraints on per-label indexes. (b) Performance impact of vector-label correlation. (c) Tree traversal overhead in \sys's search algorithm.}
  \label{fig:ablation2}
\end{figure}

We conduct three ablation studies to validate \sys's design and evaluate its robustness across different dataset characteristics.

\parab{Structural Constraints.} \sys enforces structural sharing by constraining per-label indexes to follow the base index's hierarchical partitioning, which potentially impacts performance since per-label indexes cannot freely adapt to label distributions. To evaluate this trade-off, we compare \sys against a variant where per-label indexes are constructed independently with the same hyper-parameters. Fig.~\ref{fig:ablation2}(a) shows nearly identical performance between both versions, demonstrating that structural constraints have minimal impact on the quality of embedded per-label indexes. This finding is further supported by the similar performance between \sys with predicate indexed versus per-predicate indexing in \autoref{sec:eval:query_performance}.

\parab{Vector-Label Correlation.}
Real-world datasets exhibit varying degrees of vector-label correlation—the tendency for vectors sharing the same label to cluster together in the embedding space due to semantic similarity. To assess \sys's robustness across datasets with different correlation patterns, we test on a shuffled variant of YFCC-10M where labels are randomly reassigned, effectively eliminating vector-label correlation. Fig.~\ref{fig:ablation2}(b) shows nearly identical recall-throughput trade-offs on both original and shuffled datasets, confirming \sys's robust performance across varied data distributions. We additionally quantify correlation using the "Query Correlation" metric from ACORN~\cite{patel2024acorn}, which compares a query's distance to the set of qualified vectors against a random set of equal size; larger values indicate stronger clustering by label. On YFCC-10M, the average query correlation is 0.161, whereas on the label-shuffled dataset it is 0.028, showing the effectiveness of label shuffling and confirming that the original dataset exhibits meaningful correlation.

\parab{Tree Traversal Overhead.} Unlike graph-based indexes where all distance computations target individual vectors, \sys performs distance computations against both cluster centroids (during tree traversal) and vectors (during buffer scanning). To quantify the overhead of tree traversal, we measure the proportion of distance computations used for tree traversal versus buffer scanning. Fig.~\ref{fig:ablation2}(c) shows that traversal overhead decreases from 26.9\% to 18.9\% as dataset size grows from 2M to 10M vectors, demonstrating that the overhead remains moderate and diminishes at larger scales.

\section{Related Work}
\label{sec:related}

\parab{Approximate Nearest Neighbor Search.}
As vector similarity search becomes increasingly important across various modern applications, many efficient vector indexing techniques have been proposed. Graph-based indexes~\cite{malkov2018hnsw,fu2019nsg,fu2016efanna,harwood2016fanng,iwasaki2018optimization,jayaram2019diskann} construct proximity graphs, where vectors are connected to their nearest neighbors. At query time, a best-first beam search begins from a set of seed nodes and navigates through the graph to locate vectors near the query.
Partition-based indexes, which can be further categorized into clustering-based~\cite{johnson2019faiss,babenko2014inverted,baranchuk2018revisiting,zhang2014composite,chen2021spann}, hashing-based~\cite{gionis1999similarity,andoni2015practical,xu2011complementary,shrivastava2014improved,shrivastava2014asymmetric,neyshabur2015symmetric} and tree-based~\cite{muja2009fast,annoy,muja2014scalable} approaches, partition the vector space into sub-spaces and represent vectors by the sub-space to which they belong. During search, the algorithm navigates the index to identify the sub-spaces closest to the query vector and examines all vectors within them.

\parab{Filtered ANNS.}
Existing approaches to filtered ANNS follow three main paradigms. \emph{Metadata filtering} strategies, widely adopted in production vector databases~\cite{wei2020analyticdb,wang2021milvus}, use cost models to dynamically select optimal filtering strategies based on estimated selectivity. 
\emph{Per-label indexing} constructs dedicated vector indexes for each label to eliminate query-time filtering overhead. While applying this approach naively incurs prohibitive memory overhead due to storing multiple copies of vectors, Parlay-IVF~\cite{parlayivf} addresses this limitation through shared vector storage across per-label indexes.

To further improve performance beyond these straightforward approaches, \emph{specialized indexes} tailor both index construction and search algorithms for filtered ANNS. Fused distance approaches~\cite{wang2022navigable,wu2022hqann} incorporate attribute similarity into vector distance to bias search toward qualified vectors, though they limit query expressiveness by requiring all attributes to be specified. Graph densification methods~\cite{gollapudi2023filtered,patel2024acorn} build high-degree graphs to ensure connectivity among qualifying vectors, then compress them to reduce memory usage, but face prohibitive construction costs at low selectivity. Subgraph stitching~\cite{cai2024navigating} constructs separate graphs for each unique label set and adds interconnecting edges, yet suffers from degraded index quality when these graphs are highly fragmented with few vectors in each.

Beyond categorical filtering, several works address \emph{range filtering} on numeric attributes, where queries specify value ranges rather than discrete labels. ARKGraph~\cite{zuo2023arkgraph} and SeRF~\cite{zuo2024serf} merge and compress per-range indexes to reduce storage overhead, while iRangeGraph~\cite{xu2024irangegraph} constructs graph indexes for fewer ranges and dynamically combines them at query time. DIGRA~\cite{jiang2025digra} and RangePQ~\cite{zhang2025efficient} extend these approaches with dynamic update capabilities. Unlike \sys's focus on categorical label filtering, range filtering leverages the total order of numeric attributes to optimize for contiguous value ranges, representing a complementary but distinct problem domain.

\section{Conclusion}
This paper introduced \sys, a partition-based index that complements existing graph-based approaches for low-selectivity filtered ANNS. \sys employs hierarchical clustering to build specialized per-label indexes within a shared clustering tree, adapting clustering granularity to label distributions while maintaining low overhead through structural sharing. \sys supports both single-label search and complex predicate search by dynamically constructing temporary indexes with minimal overhead. Our evaluation demonstrates that integrating \sys with state-of-the-art graph indexes reduces low-selectivity query latency by up to 20.9$\times$ compared to pre-filtering fallback, while increasing construction time and memory footprint by only 5.5\% and 4.3\%, respectively. 
\sys's source code is publicly available at \url{https://github.com/hatsu3/curator-v2}.

\begin{acks}
We thank the anonymous reviewers for their valuable feedback.
This research was partially supported by NSF AI Institute on Edge Computing (Athena, grant No. 2112562) and NSF grants (2503010, 2402696, 2238665, 2402823), and by gifts from Adobe, Amazon, Meta, and IBM.
\end{acks}

\bibliographystyle{ACM-Reference-Format}
\bibliography{99-ref}

\end{document}